\pdfoutput=1

\documentclass[12pt]{article}
\usepackage[
top = 2.5cm, 
bottom = 2.5cm,  
left = 2.5cm,  
right = 2.5cm]{geometry}

\usepackage{latexsym} 
\usepackage{verbatim}
\usepackage{tikz}
\usetikzlibrary{matrix}
\usepackage{color}
\usepackage{graphicx,amssymb,amsfonts,amsmath,amssymb,amscd,amstext, mathrsfs}
\usepackage{graphicx} 
\usepackage{bbm}
\usepackage{dsfont}
\usepackage{xcolor}
\usepackage[colorlinks=true, citecolor=black, linkcolor=black, allcolors=black]{hyperref}   
\usepackage{amsmath}
\usepackage{empheq}
\usepackage{hyperref}

\newlength\dlf

\def\be{\begin{eqnarray}}
\def\ee{\end{eqnarray}}
\def \bea {\begin{equation}}
\def \eea {\end{equation}}

\def \nn {\nonumber}

\def \si{\sigma}
\def \de{\delta}

\def \rr {\raise.35ex\hbox{\small $\prime$}\kern-.17em{\mbox{\large $\imath$}}}

\def \dels {\partial\kern-.5em / \kern.5em}
\def \As {{A\kern-.5em / \kern.5em}}
\def \Ds {D\kern-.7em / \kern.5em}

\def \k {\kappa}

\def \B {\tilde{B}}

\def\frac#1#2{{#1\over #2}}

\newcommand{\<}{\langle}
\renewcommand{\>}{\rangle}

\def \iffa {\iffalse} 

\def \ed  {\end{document}}

\def\nn{\nonumber}
\def \k {\kappa}

\def \ed  {\end{document}}

\DeclareFontShape{OT1}{cmr}{mx}{n}{<->cmr10}{}
\newcommand{\titlefont}{\fontseries{mx}\selectfont}

\def \de{\delta}
\def \si{\sigma}

\def \hA{{\widehat A}}

\def \pr{\partial}

\def \hs{{\hat s}}

\def \half{{\textstyle {1 \over 2}}}

\def \B{{\cal B}}

\def \I{{\cal I}}

\usepackage{cite}

\begin{document}

\begin{titlepage}

\begin{flushright} 
\end{flushright}

\begin{center} 
\vspace{1cm}  

{\fontsize{21pt}{0pt}{\titlefont{$d>2$ Stress-Tensor OPE near a Line}}}   
\vspace{2cm}  
\\
 {\fontsize{12pt}{0pt}{\titlefont{Kuo-Wei Huang}}}   
\\ 
\vspace{0.8cm} 
\it{
Department of Physics, 
Boston University, \\
Commonwealth Avenue, Boston, MA 02215, USA
}\\
\end{center}
\vspace{2.2cm} 

{\noindent We study the $TT$ OPE in $d>2$ CFTs whose bulk dual is Einstein gravity.
Directly from the $TT$ OPE, we obtain, in a certain null-like limit, an algebraic structure consistent with the Jacobi identity:
 $[{\cal L}_m, {\cal L}_n]= (m-n) {\cal L}_{m+n}+ C m (m^2-1) \delta_{m+n,0}$. 
The dimensionless constant $C$ is proportional to the central charge $C_T$. 
Transverse integrals in the definition of ${\cal L}_m$ play a crucial role. We comment on the corresponding limiting procedure and point out a curiosity related to the central term. 
A connection between the $d>2$ near-lightcone stress-tensor conformal block and the $d=2$ $\cal W$-algebra is observed.  
This note is motivated by the search for a field-theoretic derivation of $d>2$ correlators in strong coupling critical phenomena.
}

\end{titlepage}

\addtolength{\parskip}{0.8 ex}
\jot=1.2 ex


\subsection*{{\fontsize{16pt}{0pt}{\titlefont{1. Introduction}}}}

\setcounter{page}{2}

The $TT$ OPE in $d=2$ CFT
\begin{align}
\label{2dTT}
T(z_1) T(z_2) ={ {\bf c} \over 2 s^4} + {2\over s^2} T(z_2)
+ {1\over s} \partial_{z_2} T (z_2)  + {\cal O}(\partial^2 T) \ ,  ~~~s=z_1-z_2  
\end{align} 
 leads to the Virasoro algebra:
\begin{align}
\label{V}
[L_m, L_n] = (m-n) L_{m+n} +   {  {\bf c} \over 12}  m (m^2-1) \delta_{m+n,0}  \ ,~~~~ L_m = {1\over 2 \pi i}\oint dz ~z^{m+1}~T(z) \ .
\end{align}
The Virasoro algebra is omnipresent in two-dimensional critical phenomena \cite{BELAVIN1984333} and has enormous implications; 
in particular, the algebra provides a non-perturbative derivation of $d=2$ conformal correlators.  
In higher dimensions, the general $TT$ OPE is contaminated by many model-dependent details. However, we ask the question:
can one generalize the derivation \eqref{2dTT}-\eqref{V} to $d>2$ CFT in certain physical limits?  

Over 27 years ago, Osborn and Petkos \cite{Osborn:1993cr} computed the stress-tensor contribution to the $d>2$ $TT$ OPE, but we have not found any computation based on such an explicit $TT$ OPE. A reason, presumably, is that the $TT$ OPE is complicated. Given the recent developments of gauge/gravity correspondence and $d>2$ strongly coupled field theories, we find it necessary to revisit the $d>2$ $TT$ OPE structure. In this note, we adopt the following two simplifying limits to reduce the complexity of the $TT$ OPE:  
\\
(i) Infinitely large higher-spin gap; 
(ii) Null/lightcone-like limit.\footnote{~We use ``-like"  to distinguish our limiting procedure from similar limits used in the literature: the null-line limit is often defined by directly setting $x^+ = x^2_\perp =0$ in the Lorentzian signature, where $x^\pm= x^0\pm x^1$ and $x_\perp$ denotes transverse directions.}

As shown in \cite{Camanho:2014apa, Heemskerk:2009pn, Afkhami-Jeddi:2016ntf}, the gap $\Delta_{\rm gap}$ to the lightest spin $>2$ single-trace primary controls the higher-order corrections to Einstein gravity; the limit $\Delta_{\rm gap} \to \infty$ then selects CFTs with an Einstein gravity bulk dual.  We will focus on stress-tensor contribution to the $TT$ OPE and suppress other primary operators.\footnote{~For instance, the three-point function $\langle T {\cal O}{\cal O} \rangle$ is suppressed as $\Delta_{\rm gap} \to \infty$ \cite{Meltzer:2017rtf}.} On the other hand, the lightcone limit has been adopted in the recent computation of the multi-stress-tensor OPE data in $d>2$ holographic CFTs \cite{Fitzpatrick:2019zqz, Fitzpatrick:2019efk, Fitzpatrick:2020yjb, Li:2019tpf, Kulaxizi:2019tkd,  Karlsson:2019dbd, Karlsson:2020ghx, Li:2019zba,  Li:2020dqm, Karlsson:2019txu, Parnachev:2020fna, Parnachev:2020zbr, Karlsson:2021duj}. The near-lightcone correlator at large central charge $C_T$ is independent of higher-curvature terms in the purely gravitational action \cite{Fitzpatrick:2019zqz}; however, the correlator depends on certain non-minimal coupling bulk interactions which are suppressed at an infinite gap \cite{Fitzpatrick:2020yjb}. These results suggest that the simplest starting point is to impose the limits (i), (ii) on the $TT$ OPE.\footnote{~We should also assume the usual large $N$ and large $C_T$ limits but we will keep  $C_T$ in our expressions.} 

In general, the results depend on the order of limits (i.e. limiting procedure). A related motivation of this work is to help identify a lightcone-like limiting procedure that may be implemented to compute multi-stress-tensor OPE coefficients and near-lightcone correlators in $d>2$ CFTs from the first principle via a Virasoro-like field-theoretic approach. 

The main result of this work is that we find a  structure similar to \eqref{V} in higher dimensions. Intuitively, one may expect that a Virasoro-like structure arises because the null-like limit brings stress tensors close to a line, a picture reminiscent of the two-dimensional case where $T =-  {\pi\over 2} T^{\bar z \bar z}(z)$, $\bar T = -  {\pi\over 2} T^{zz}(\bar z)$ are holomorphic and anti-holomorphic functions, respectively. Note we are not introducing a physical line-defect.  

This paper is organized as follows. 
In Section 2, we discuss the stress-tensor OPE structure. Some detailed expressions can be found in Appendix.  
We focus on $d=4$ for concreteness and expect that our results generalize to other dimensions. In Section 3, we consider a null-line limit and obtain a Virasoro-like commutator via  the stress-tensor OPE.  
 A $d=4$ single-stress-tensor-exchange derivation without explicitly using an algebra is discussed in Section 4, where we point out a curiosity related to the central term.  
We observe a connection between the $d>2$ lightcone stress-tensor conformal block and the central-term of the $d=2$ ${\cal W}$-algebra.  

\subsection*{{\fontsize{16pt}{0pt}{\titlefont{2. Stress-tensor OPE}}}}

 Our starting point is the stress-tensor contribution to the $d=4$ $TT$ OPE \cite{Osborn:1993cr}: 
\be
\label{TT4d}
T^{\mu\nu}(x_1) T^{\si\rho}(x_2) = C_T {\I^{\mu\nu,\si\rho}(s)\over s^{8}}
+ \hA^{\mu\nu\si\rho}_{~~~~~\alpha\beta}(s) T^{\alpha\beta}(x_2) +
B^{\mu\nu\si\rho}_{~~~~~\alpha\beta\lambda}(s) \pr^\lambda T^{\alpha\beta}(x_2)  + {\cal O}(\partial^2 T) 
\ee
where $s=x_1-x_2$. The first term has the familiar form:
\begin{align}
{\cal I}^{\mu\nu,\sigma \rho}(s) = {1\over 2} \Big( { I}^{\mu\sigma}(s) { I}^{\nu\rho}(s)+ { I}^{\mu\rho}(s) { I}^{\nu\sigma}(s)\Big)-{1\over 4} \delta^{\mu\nu} \delta^{\sigma \rho} \ , ~~~
 { I}^{\mu\sigma}(s) = \delta^{\mu\sigma}- 2 {s^\mu s^\sigma \over s^2} \ .
\end{align}
The structures of $\hA_{\mu\nu\si\rho\alpha\beta}$ and $B_{\mu\nu\si\rho\alpha\beta\lambda}$ are cumbersome so we put them in the appendix.  
As noted in \cite{Osborn:1993cr}, there are three undetermined coefficients in the $TT$ OPE, denoted as $a,b,c$. 
The central charge $C_T$ is given by\footnote{The parameter $c$ here should not be confused with the central charge in two dimensions where $C_T= {{\bf c}\over 2 \pi^2}$. } 
\begin{align}
C_T= {\pi^2\over 3} (14a-2b-5c) \ .
\end{align} 

In the lightcone limit, the relevant contribution is the lightcone component of the stress tensor, $T^{++}$.  We will mostly work in the Euclidean space and adopt the line element $ds^2= dz d\bar z +  \sum_{i=1,2} (d {x_\perp^{(i)}})^2$ where $z, \bar z$, the Euclidean analogue of the lightcone coordinates, are complex coordinates. 
 We will then focus on the $T^{zz}T^{zz}$ OPE. 

The $TT$ OPE simplifies significantly when one focuses on the $T^{zz}$ component.  Using \eqref{TT4d}, we obtain 
\be
T^{zz}(x_1) T^{zz}(x_2) =   C_T {\I^{zz,zz}(s)\over s^{8}}+  \hA^{ z z  z z}_{~~~~zz}(s) T^{zz}(x_2) +
B^{ z  z  z z}_{~~~~zz\lambda}(s) \pr^\lambda T^{zz}(x_2)  + {\cal O}(\partial^2 T)  
\ee
where $ {\I^{zz,zz}}=  {4 {(s^z)}^4 \over s^{4}}$,  
\begin{align}
\label{Az}
\hA^{ z z  z z}_{~~~~zz}&= { 4 (s^z)^2 \over  C_T s^{10}}  \Big( (2 b + c)(s^\perp)^4 -2 s^z s^{\bar z}\big( (8 a - b - 3 c)  (s^\perp)^2+(6 a - b - 2 c) s^z s^{\bar z} \big) \Big) \ , \\
B^{ z  z  z z}_{~~~~zzz} &=  {s^-\over 4} \hA^{ z z  z z}_{~~~~zz} \ , \\
B^{zzzz}_ {~~~~zz\perp} &=  {s^\perp \over 2} \hA^{z z z  z}_{~~~~zz}  \  , \\
B^{zzzz}_{~~~~zz \bar z} &= { (s^{z})^3 \over 9 C_T s^{10}}  \Big((64 a + 18 b - 11 c)(s^\perp)^4  \nn\\
&~~~~~~~~~~~~~~~~~~~~~~~~~~~ -2 s^z s^{\bar z} \big( 4 (3 a - b - 2 c) (s^\perp)^2+(26 a - 4 b - 9 c) s^z s^{\bar z} \big) \Big)  \ .
\end{align}
We will argue that higher-order pieces, ${\cal O}(\partial^2 T)$, are irrelevant when imposing the null-like limit considered in Sec. 3. 
Observe that, from \eqref{Az}, $(8 a - b - 3 c) +(6 a - b - 2 c)= {3\over \pi^2} C_T$. 
While this combination is interesting, we here consider a large $N$, large-gap condition which places strong constraints on the flux parameters ``$t_2$" and ``$t_4$" of the energy flux escaping to null infinity \cite{Hofman:2008ar, Henningson:1998gx, Belin:2019mnx}:
\begin{align}
\label{LG}
t_2 = { 30 (13 a+4b-3c) \over  (14a-2b-5c)} = 0 \ , ~~~~~~t_4 = {-15 (81 a+32b-20c) \over  2 (14a-2b-5c)} = 0 \ .
\end{align}  
It is worth mentioning that two trace-anomaly central charges become the same under these conditions.  By imposing $t_2=t_4=0$ without first requiring a strictly infinite $C_T$, we can reduce three parameters to one parameter.

\subsection*{{\fontsize{16pt}{0pt}{\titlefont{3. Stress-tensor OPE near a line and a Virasoro-like commutator}}}}

Consider the following operator in $d=4$:  
\begin{align}
{\cal L}_m= {\k\over 2 \pi i} \oint d\bar z~ \bar z^{m+1}  \int d^2 x_\perp ~ T^{zz}(z, \bar z, x_\perp^{(1)}, x_\perp^{(2)}) \ ,~~~  \int d^2 x_\perp = \int_0^l d x_\perp^{(1)} \int_0^l dx_\perp^{(2)}
\end{align} 
in the null-like limit ${z \to 0, l\to 0}$.\footnote{~One may perform a Wick-rotation to Lorentzian space and impose the lightcone limit, and then Wick-rotate back to the Euclidean space to carry out the $\bar z$ integral via the residue theorem.  One may also formally impose a small $z$ limit directly in Euclidean space, which is what we will do here.  For two stress tensors, we take a small $s^z$.   A similar analysis applies to the $T^{\bar z \bar z}T^{\bar z \bar z}$ OPE if one instead chooses  a small $s^{\bar z}$ limit.}   We will determine the overall normalization factor  $\k$ later. The interpretation of the small $l$ limit  is that we consider the stress-tensor contribution near a two-dimensional plane. We are interested in computing  the commutator $[{\cal L}_m,{\cal L}_n]$.  The transverse  integrals are crucial, as we will see, for extracting a central extension consistent with a Witt-like algebra.\footnote{~This construction is essentially the same as the mode operator introduced in \cite{Huang:2020ycs}, but in that work the author adopts a different limiting procedure.  See also \cite{Casini:2017roe, Huang:2019fog, Besken:2020snx, Belin:2020lsr} for related discussions.} 
 
Let us first consider the $c$-number term which is controlled by the stress-tensor two-point function.  
After performing the transverse integrations, we consider a small $s^z$ expansion: 
\be
\label{ctermint}
\lim_{s^z\to \delta} \int d^4 x_\perp {C_T \I^{zz,zz}(s)\over s^{8}} = \frac{4 \pi  C_T}{5  (s^{\bar z})^5} {l^2 \over \delta}
- {7 \pi  C_T \over 16 (s^{\bar z})^{9/2}} {l \over \sqrt \delta} +\frac{C_T}{5 (s^{\bar z})^4} + \frac{(356+315 \pi ) C_T }{14400}  {\delta ^4\over  l^8} + \cdots
\ee
We would like to extract the cutoff-independent piece.  We do so by next imposing a $l \to 0$ limit such that the first two terms are suppressed. The last piece of \eqref{ctermint} and higher-order terms, although divergent as $l \to 0$, do not have a $\bar z$-pole and thus do not contribute to the commutator.  The ${C_T \over (s^{\bar z})^4}$ term 
shares the same form as the $c$-number term in the $d=2$ $TT$ OPE \eqref{2dTT}.   The transverse integrals compensate for the additional dimensions of the $d>2$ $TT$ OPE. This $c$-number-term derivation does not require a large-gap condition.   The Cauchy's integral formula now leads to\footnote{~In general $d$, we find 
\begin{align}
[{\cal L}_m,{\cal L}_n]|_{C_T}= (-1)^{d}  \k^2 {4 C_T\over \Gamma(d+2)} m (m^2-1) \delta_{m+n,0} \ . \nn
\end{align}} 
\begin{align}
\label{fC}
[{\cal L}_m,{\cal L}_n]|_{C_T}=  ({\k\over 2 \pi i})^2\oint_{{\cal C}(0)}  d\bar z_2 \bar z_2^{n+1} \oint_{{\cal C}(\bar z_2)} d\bar z_1  \bar z_1^{m+1}   \frac{C_T}{5 (s^{\bar z})^4} = \k^2 { C_T\over 30} m (m^2-1) \delta_{m+n,0} \ .
\end{align} 

We next turn to the operator part of the $TT$ OPE,  keeping explicit $a,b,c$ parameters and imposing the conditions \eqref{LG} at the end. We evaluate 
\begin{align}
\label{As}
\lim_{s^z\to \delta} \int d^2 (x_1)_\perp \Big( \hA^{zzzz}_{~~~~zz}(x_1-x_2) T^{zz}(x_2) \Big) &=  f(a,b,c)  {T^{zz}(x_2)\over  \pi (s^{\bar z})^2} + {\cal O} (\delta)  
\end{align} 
where $f(a,b,c)={ -52 a+10 b+19 c\over 14 a-2 b-5 c}$.  The leading-order term is cutoff-independent and only depends on $(x_2)_\perp$ through the stress tensor. To take the small $s^z$ limit, we may assume $T^{zz}(x_2)$ is a suitable test function having a finite contribution only near $y_2=z_2 =0$, and then perform all the transverse  integrations before imposing the small $s^z$ limit. But we find it simpler, as we did above, to take $s^z \to \delta$ right after performing the integrations over the first set of transverse coordinates $(x_1)_\perp$.\footnote{~In the process of simplifying \eqref{As},  we formally assume $l>(x_2)_{\perp}>0$ to adopt the identity $\tan^{-1}(X)+\tan^{-1}(1/X)= \pi/2$ with $X>0$. This, strictly speaking, means the end points of the $(x_2)_{\perp}$ integrals should be removed.
}
 It is now straightforward to complete the rest of the integrations:
\begin{align}
[{\cal L}_m,{\cal L}_n]|_{A\rm{-term}}&=  ({\k\over 2 \pi i})^2 {f(a,b,c)\over \pi} \oint_{{\cal C}(0)}  d\bar z_2 \bar z_2^{n+1} \oint_{{\cal C}(\bar z_2)} d\bar z_1  \bar z_1^{m+1}  \int d^2 (x_2)_\perp  {  T^{zz}(x_2)\over (s^{\bar z})^2}   \nn\\
\label{fA}
&=  {\k f(a,b,c)\over \pi}  ~  (m+1)  {\cal L}_{m+n} \ .
\end{align}  
For the second-order term in the $TT$ OPE, we have  
\begin{align}
\lim_{s^z \to \delta} \int d^2 (x_1)_\perp \Big( B^{zzzz}_{~~~~zz z}(s) \pr^z T^{zz}(x_2) \Big) 
&= 2 \lim_{s^z \to \delta} \int d^2 (x_1)_\perp \Big( B^{zzzz}_{~~~~ zz z}(s) \pr_{\bar z} T^{zz}(x_2) \Big)  \nn\\
\label{iBdT}
&= f(a,b,c)  {\pr_{\bar z}  T^{zz}(x_2)\over 2 \pi s^{\bar z}} + {\cal O} (\delta) \ ,  \\
\lim_{s^z \to \delta} \int d^2 (x_1)_\perp \Big(  B^{z z z z}_{~~~~zz\perp}(s) \pr_\perp T^{zz}(x_2) \Big)&= {\cal O} (\delta)  \ . 
\end{align} 
Since we only focus on the $T^{zz}$ component in the null-like limit and effectively turn off other components of the stress tensor, the conservation of the stress tensor implies that we also drop $\pr_z T^{zz}$.   Observe that the structures (including the relative coefficients) of \eqref{As}, \eqref{iBdT} are the same as the two-dimensional case \eqref{2dTT}. 
From \eqref{iBdT}, we get 
\begin{align}
\label{fB}
[{\cal L}_m,{\cal L}_n]|_{B\rm{-term}}=  - {\k f(a,b,c)\over 2 \pi}  ~  (m+n+2)  {\cal L}_{m+n} \ .
\end{align}  
Similar to the corresponding $d=2$ computation, we have performed an integration by parts to evaluate the $\pr_{\bar z} T^{zz}$ term. 

We will not include the higher-order corrections ${\cal O}(\partial^2 T)$ in the $TT$ OPE, but, based on the pattern \eqref{As}, \eqref{iBdT}, it seems 
reasonable to assume that the higher-order terms do not have a relevant pole in the null-like limit.  

Combining \eqref{fB}, \eqref{fA}, and \eqref{fC}, the result is 
\begin{align}
\label{4dV}
[{\cal L}_m,{\cal L}_n]=  (m-n) {\cal L}_{m+n} +  \k^2 {C_T\over 30} m (m^2-1) \delta_{m+n,0}  ~~~\ ,  ~~~~~ \k= {2 \pi \over f(a,b,c)}  \ .
\end{align} 

We choose a normalization $\k$ such that the non-central term has a simple coefficient.  
If we now impose the conditions listed in \eqref{LG}, we find the normalization factor to be $\k= - {90 \pi \over 181}$.\footnote{~An overall rescaling of the mode operator ${\cal L}_m$ should not affect a scalar correlator computation. But one might wonder if the ``right" proportionality constant should instead be $\k= - {90 \pi \over 180} =- {\pi \over 2}$.   If we formally adopt free-theory values of $a, b, c$ \cite{Osborn:1993cr}, we notice that $\k= - {\pi \over 2}$ for both a fermion and a $U(1)$ gauge field, but $\k= - {18 \pi \over 37}$ for a scalar.  In fact, $\k=- {\pi \over 2}$ is true only under the condition $4a+2b-c=0$, which holds for both a free fermion and a $U(1)$ gauge field, but a free scalar has $4a+2b-c= -{1\over 9 \pi^6}$.  (In $d=2$, on the other hand, $4a+2b-c=0$ for both a free scalar and a free fermion.)}

To summarize, we have described a null-line-like limiting procedure that allows us to extract an algebraic structure from the $TT$ OPE. The result \eqref{4dV} is strikingly similar to the two-dimensional Virasoro algebra. We do not use holographic duality here, but it would be nice to find a potential connection to the AdS/CFT computation discussed some time ago \cite{Banados:1999tw, Brecher:2000pa} where a higher-dimensional generalization to the Brown-Henneaux symmetry \cite{Brown:1986nw} was identified in a certain infinite momentum frame. Most likely, whether or not there is a Virosoro-like structure at infinity depends on boundary conditions.\footnote{~I thank Gary Gibbons for related remarks.}

 It would certainly be of great interest to extend the two-dimensional CFT analysis to higher dimensions in the null/lightcone-like limit, where one expects to find relatively robust structures.  By first focusing on a special class of higher-dimensional CFTs with an Einstein gravity dual, we would like to know if there is an effective algebraic derivation of the multi-stress-tensor OPE coefficients and conformal correlators. Considering perturbative corrections due to a large but finite higher-spin gap could be interesting as well.

\subsection*{{\fontsize{16pt}{0pt}{\titlefont{4. A $d=4$ single-stress-tensor-exchange derivation}}}}

Let us conclude this note by presenting some observations, which hopefully shed light on more general cases.  In the following, we point out a simple derivation of the $d=4$ near-lightcone conformal scalar correlator via a mode summation.\footnote{~The derivation presented here is simpler than previous work \cite{Huang:2020ycs} and we can avoid an arbitrary parameter introduced in that paper.}  
This derivation does {\it not} explicitly rely on an algebra.  In fact, as we will see, this derivation presents a central-term curiosity. 

The scalar four-point conformal correlator can be written in terms of the conformal block decomposition\cite{Dolan:2000ut}: 
\begin{align}
\label{CB}
\langle  {\cal O}_H (\infty) {\cal O}_H(1) {\cal O}_L(z,\bar z) {\cal O}_L(0) \rangle
= \sum_{\Delta_T, J}  c_{\rm OPE} (\Delta_T, J) {B(z, \bar z, \tau, J)\over (z \bar z)^{\Delta_L}}  
\end{align} 
where the twist of an operator is its dimension minus its spin, $\tau=\Delta_T-J$.
We formally name ${\cal O}_H$ the ``heavy"  scalar  and  ${\cal O}_L$ the ``light" scalar although the heavy-light limit (i.e. $\Delta_H, C_T \to \infty$ with $\Delta_H/C_T$ fixed and $\Delta_L \sim {\cal O}(1)$) does not play a special role in the single-stress-tensor-exchange computation. We adopt this notation as an example which is useful to compare with the literature that discusses multi-stress-tensor contributions to the heavy-light correlator. 
 The Ward identity fixes the stress-tensor OPE coefficient to be $c_{\rm OPE} (4, 2)= {\Delta_H \Delta_L\over 9 \pi^4 C_T}$ in the convention of \eqref{TT4d}.
The conformal block is given by\footnote{Our convention differs by an overall factor of $ (-{1\over 2})^J$ from the convention used in Dolan and Osborn \cite{Dolan:2000ut}.}  
\begin{align}
\label{B}
&B(z, \bar z, \tau, J)\\
&= {z \bar z \over z-\bar z}
\Big[z^{\tau + 2J\over 2} {\bar z}^{\tau-2 \over 2} \, _2F_1\Big({\tau + 2J \over 2},{\tau + 2J \over 2}; \tau + 2J;  z\Big)
\, _2F_1\Big({\tau-2 \over 2},{\tau-2 \over 2}; \tau-2;  \bar z\Big) - (z \leftrightarrow \bar z) \Big]\ . \nn
\end{align} 
 In the limit $z\to 0$, the stress-tensor contribution in $d=4$ reads\footnote{~One can also choose $\bar z \to 0$ as the lightcone limit.} 
\begin{align}
\label{Tex}
  \lim_{z\to 0} & \Big( (z \bar z)^{\Delta_L} \langle  {\cal O}_H (\infty) {\cal O}_H(1) {\cal O}_L(z,\bar z) {\cal O}_L(0) \rangle|_{T} \Big)\nn\\
& =  {1 \over 9  \pi^4} {\Delta_H \Delta_L \over C_T} \bar z^3 ~ {}_2 F_1(3,3,6, \bar z)  z  + {\cal O}( z^2) \\
& = {10 \over 3 \pi^4} {\Delta_H \Delta_L \over C_T}{3 ( \bar z-2)  \bar z- \big(6 +( \bar z-6) \bar z \big) \ln(1- \bar z) \over \bar z^2}z  + {\cal O}( z^2)\ . \nn
\end{align} 
 The higher-order pieces represent multi-stress-tensor contributions to the correlator.  

 It is instructive if we temporarily forget about the algebra and instead adopt the following operator:
\begin{align}
\label{lr}
\widetilde {\cal L}_m= \lim_{z_T\to \delta } \oint {d \bar z_T\over 2 \pi i}~ \bar z_T^{m+2}~ T^{zz} (z_T, \bar z_T, x^\perp=0) \ .
\end{align}  
Notice we directly set $x^\perp=0$ in this definition.  The $z_T\to \delta$ represents the null-line limit for the stress tensor.  
The notation ``$m+2$" will result in slightly more symmetric expressions in the following computation. 
The mode operator \eqref{lr} is essentially the same as the lightray operator  which does  not contain transverse   integrals \cite{Casini:2017roe, Kravchuk:2018htv, Besken:2020snx, Belin:2020lsr}. We here use the Euclidean signature with complex coordinates $z, \bar z $.  
Similar to the $d=2$ case, we may expect that the stress-tensor-exchange contribution can be computed via the following mode summation:
\begin{align}
\label{V1}
{\cal V}_T = \lim_{z  \to 0} \sum^\infty_{m=m^*} {\<{\cal O}_H(\infty) {\cal O}_H(1) \widetilde {\cal L}^\dagger_{m} \>\<\widetilde {\cal L}_{m}{\cal O}_L (z, \bar z ){\cal O}_L(0)  \> \over   
 ~\<{\cal O}_H(\infty) {\cal O}_H(1)  \> {\cal N}_m  \<{\cal O}_L (z, \bar z ){\cal O}_L(0)  \> } 
\end{align}
where the normalization factor is ${\cal N}_m= \<\widetilde {\cal L}_{m}\widetilde {\cal L}^\dagger_{m}\>$. We will find $m^*=3$.

Using the three-point function 
\begin{align}
\label{TOO}
\< T^{\mu\nu}(x_1) {\cal O}_{\Delta} (x_2) {\cal O}_{\Delta}(x_3) \> 
= {c_{T{\cal O}{\cal O}}\over x^4_{12} x^4_{13} x^{2\Delta-4}_{23}} \Big({X^\mu X^\nu \over X^2} -{\delta^{\mu\nu}\over 4} \Big) 
\end{align}
with $X^\mu = { x^\mu_{12}/x^2_{12} }-{ x^\mu_{13}/ x^2_{13} }$ and $c_{T{\cal O}{\cal O}}= - {2 \Delta\over 3 \pi^2}$, we first obtain  
\begin{align}
\lim_{z_T\to \delta } \lim_{z\to 0} {\<{\cal \widetilde  L}_{m}{\cal O}_L (z, \bar z ){\cal O}_L(0)  \> \over\<{\cal O}_L (z, \bar z ){\cal O}_L(0)  \>}
&=  - {2 \Delta_L \over 3 \pi^2} \lim_{z_T\to \delta } \lim_{z\to 0}  \oint_{{\cal C}(\bar z)} {d \bar z_T\over 2 \pi i}~ \bar z_T^{m+2}~ \frac{\bar z^3 z}{  {\bar z_T^3 z_T (\bar z_T-\bar z )^3 ( z_T- z )} } \nn\\
&= - {2 \Delta_L \over 3 \pi^2} \lim_{z_T\to \delta }  \lim_{z\to 0} \frac{(m-1) (m-2)   \bar z^m }{2 z_T (z_T-z)}z  \nn\\
&= - {\Delta_L \over 3 \pi^2} \frac{(m-1) (m-2) \bar z^{m}}{ \delta^2}z\ , 
\end{align} 
where we introduce a short-distance cutoff $\delta$. We shall find that the final four-point scalar correlator is independent of the UV cutoff.  
It is important to adopt a proper order of limits. 

On the other hand, by taking ${\cal \widetilde  L}^\dagger_{m}= {\cal \widetilde  L}_{-m}$, we find 
\begin{align}
{\<{\cal O}_H(\infty) {\cal O}_H(1) {\cal \widetilde L}^\dagger_{m} \> \over   ~\<{\cal O}_H(\infty) {\cal O}_H(1)  \>   }  
&=  {2 \Delta_H \over 3 \pi^2} \lim_{z_T\to \delta }  \oint_{{\cal C}(1)} {d \bar z_T\over 2 \pi i}~  \frac{(\bar z_T)^{-m+2} }{({\bar z_T}-1)^3 (z_T-1)}\nn\\
&= { 2\Delta_H \over 3 \pi^2} \lim_{z_T\to \delta }  \frac{ (m-1)(m-2) }{2 (z_T-1)}\nn\\
&= - { \Delta_H \over 3 \pi^2} (m-1) (m-2)   \ . 
\end{align} 
The normalization factor can be computed using the stress-tensor two-point function:
\begin{align}
{\cal N}_m= \< {\cal\widetilde L}_{m} {\cal\widetilde L}^\dagger_{m}\>
&= \lim_{s^z \to \delta }\oint_{{\cal C} (0)} {d \bar  z_{2T} \over 2 \pi i} \oint_{{\cal C}(\bar  z_{2T}) } { d \bar  z_{1T}\over 2 \pi i} ~ (\bar  z_{1T})^{m+2} ~(\bar  z_{2T})^{-m+2}   \< T^{zz}(z_{1T},\bar z_{1T}) T^{zz}(z_{2T},\bar z_{2T})  \>\nn\\
&=  \lim_{s^z \to \delta } \oint_{{\cal C} (0)} {d \bar  z_{2T} \over 2 \pi i} \oint_{{\cal C}(\bar  z_{2T}) } {d \bar  z_{1T}\over 2 \pi i} ~{ 4 C_T (\bar  z_{1T})^{m+2} ~(\bar  z_{2T})^{-m+2} \over  (\bar  z_{1T}-\bar  z_{2T})^6 (z_{1T}-  z_{2T})^2} \nn\\ 
&= {C_T\over 30} {(m+2) (m+1) m (m-1) (m-2)\over \delta^2}  \ .
\end{align} 
The UV-cutoff dependencies cancel out in the final mode summation and we obtain exactly the $d=4$ stress-tensor-exchange structure \eqref{Tex}:
\begin{align}
\label{modec}
{\cal V}_T =  { 10 \over 3 \pi^4}{ \Delta_H  \Delta_L  \over  C_T  } \sum_{m=3}^\infty {  (m-1) (m-2) \bar z^{m}  \over   m (m+1) (m+2) } z =  {1 \over 9  \pi^4} {\Delta_H \Delta_L \over C_T} \bar z^3 ~ {}_2 F_1(3,3,6,\bar z)  z \ .
\end{align}    
This computation does not require a large gap. 

It is peculiar that we are able to reproduce the $d=4$ near-lightcone correlator, including the correct OPE coefficient, via a mode summation. 
The final result is finite and cutoff-independent. Although the above single-stress-tensor computation does not rely on knowing an algebra, we would like to ask why such a derivation exists.  
Recall that, in two-dimensions, a similar derivation exists because of the Virasoro symmetry. 
Given the above $d>2$ computation, one may speculate that a certain symmetry emerges near the lightcone. 
An underlining algebra would provide a precise interpretation of the modes counting in \eqref{modec}.  
Since we have extracted a Virasoro-like commutator from the stress-tensor OPE \eqref{4dV}, it seems natural to link the correlator computation to the algebra.  

However, we find a curiosity related to the central term, or more generally, to the limiting procedure.
In the above correlator computation, we emphasize that {\it we take $x^\perp \to 0$ before imposing $z \to 0$}.\footnote{~Using this order of limits, one can include transverse integrals but the correlator result is unchanged.} The resulting ``central" term has the following $m$-dependence (in the notation of $\widetilde {\cal L}_m \sim \oint d \bar z ~\bar z^{m+2} T$):
\begin{align} 
\label{typeA}
{\rm Type~ A:}~~~~~ C_T~ (m+2) (m+1) m (m-1) (m-2) \delta_{m+n,0} \ .
\end{align} 
Such an $m$-dependence is quite different from the central term in the Virasoro-like commutator, which has the structure:  
\begin{align} 
\label{typeB}
{\rm Type~ B:}~~~~~ C_T~ (m+1) m (m-1) \delta_{m+n,0}\ . ~~~~~~~~~~~~~~~~~~~~
\end{align} 
As shown above, in this case, {\it we take $z \to 0$ before imposing $x^\perp \to 0$}. 
The Type-B central-term has the familiar form fixed by the Jacobi identity, but the correlator derivation suggests that the Type-A structure plays a non-trivial role in recovering the scalar correlator.  The Type-A structure, however, is incompatible with the Witt algebra. 

Ideally, we would like to also compute $[\widetilde {\cal L}_m, \widetilde {\cal L}_n]$ via the $d=4$ $TT$ OPE, but we find that the commutator computation using $\widetilde {\cal L}_m$ requires a higher-order term in the $TT$ OPE.  Such a computation will not be included in this note. 

On the other hand, we observe that, up to an overall coefficient, the $d=4$ Type-A structure \eqref{typeA} is identical to the central term of the ${\cal W}_3$ algebra in $d=2$ CFTs \cite{Z1985} (see \cite{Bouwknegt:1992wg} for a review): 
\begin{align} 
[W_m, W_n]
&=   \frac{{\bf c}}{360} \,  (m+2) (m+1) m (m-1) (m-2)  \delta_{m+n,0} \nn\\
& ~~~ + (m-n) \Big(\frac{1}{15} (m+n+3)(m+n+2) - \frac{1}{6} (m+2)(n+2)
\Big) L_{m+n} \nn\\
& ~~~ +  {16 \over 22+5{\bf c}} (m-n) \Lambda_{m+n} \ , \\
[L_m ,W_n]&=(2m-n) W_{m+n} \ , 
\end{align}  where  $\Lambda_m = \sum_n (L_{m-n} L_n) - \frac{3}{10} (m+3) (m+2) L_m$.  $W_m$ is the Laurent modes of a spin-3 primary current.  Note that closure of the ${\cal W}_3$ algebra requires first knowing the operator $L_m$ that satisfies the Virasoro algebra. In general (even) $d$, we find the Type-A structure is $\sim C_T ~m (m^2-1)(m^2-4)\cdots (m^2- ({d\over 2})^2)$ in the notation of $\widetilde {\cal L}_m \sim \oint d \bar z ~\bar z^{(m+{d\over 2})} T$.  

To our knowledge, a connection between $d>2$ CFT correlators and the ${\cal W}$(-like) symmetry has not been mentioned before.  
 This central-term curiosity needs to be better understood.
Perhaps exploring more general structures involving multi-stress-tensor exchanges in $d>2$ CFTs can help clarify its algebraic underpinnings.  

\begin{center}
\subsubsection*{{\fontsize{12pt}{0pt}{\titlefont{\bf {Acknowledgments}}}}} 
\end{center}

I am grateful to L. Fitzpatrick, J. Maldacena, H. Osborn, and A. Parnachev for helpful comments. 
This work was supported in part by the U.S. Department of Energy Office of Science No. DE-SC0015845 and 
the Simons Collaboration on the Nonperturbative Bootstrap. 

\newpage
\subsection*{{\fontsize{18pt}{0pt}{\titlefont{Appendix}}}}

For convenience, here we collect structures appearing in the $TT$ OPE in general $d$ \cite{Osborn:1993cr}.

\addtolength{\parskip}{0.8 ex}
\jot=0.4 ex

Define 
\begin{align}
\hA_{\mu\nu\si\rho\alpha\beta}^{\vphantom h} (s)C_T =&~ {d-2\over d+2}(4a+2b-c) H^1_{\alpha\beta\mu\nu\si\rho}(s) + {1\over d}(da+b-c) H^2_{\alpha\beta\mu\nu\si\rho}(s) \nn\\
& - {d(d-2)a-(d-2)b-2c\over d(d+2)}\bigl(H^2_{\mu\nu\si\rho\alpha\beta}(s) + H^2_{\si\rho\mu\nu\alpha\beta}(s)\bigl ) )  \nn\\
& + {2da+2b-c\over d(d-2)}H^3_{\alpha\beta\mu\nu\si\rho}(s)  - {2(d-2)a-b-c\over d(d-2)}H^4_{\alpha\beta\mu\nu\si\rho}(s) \nn\\
& - 2{(d-2)a-c\over d(d-2)}\bigl(H^3_{\mu\nu\si\rho\alpha\beta}(s) + H^3_{\si\rho\mu\nu\alpha\beta}(s)\bigl ) \nn\\
& +{(d-2)(2a+b)-dc\over d(d^2-4)}\bigl(H^4_{\mu\nu\si\rho\alpha\beta}(s) + H^4_{\si\rho\mu\nu\alpha\beta}(s)\bigl )\nn\\
& + \bigl (C\, h^5_{\mu\nu\si\rho\alpha\beta} + D (\de_{\mu\nu}^{\vphantom h}h^3_{\si\rho\alpha\beta}+\de_{\si\rho}^{\vphantom h} h^3_{\mu\nu\alpha\beta})
\bigl ) S_d \de^d(s)   ~~~~\big(S_d= {2 \pi^{d\over 2} \over \Gamma({d\over 2})}\big)
\end{align} 
where 
\begin{align}
H^1_{\mu\nu\si\rho\alpha\beta} (s) =&~\Bigl ( \pr_\mu^{\vphantom h}
\pr_\nu^{\vphantom h} - {1\over d}\, \de_{\mu \nu}^{\vphantom h}
\pr^2 \Bigl ) \, {1\over s^{d-2}} h^1_{\si\rho}(\hs)h^1_{\alpha\beta}(\hs) \\
H^2_{\mu\nu\si\rho\alpha\beta} (s) =&~\Bigl ( \de_{\si\alpha}^{\vphantom h}
\pr_\rho^{\vphantom h} \pr_\beta^{\vphantom h} + (\si\leftrightarrow \rho,
\alpha \leftrightarrow \beta) \nn\\
& - {4\over d}\, \de_{\si \rho}^{\vphantom h}
\pr_\alpha^{\vphantom h}\pr_\beta^{\vphantom h}
- {4\over d}\, \de_{\alpha \beta}^{\vphantom h}
\pr_\si^{\vphantom h}\pr_\rho^{\vphantom h}
+ {4\over d^2}\, \de_{\si \rho}^{\vphantom h}\de_{\alpha \beta}^{\vphantom h}
\pr^2 \Bigl ) \, {1\over s^{d-2}} h^1_{\mu\nu}(\hs) \\
H^3_{\mu\nu\si\rho\alpha\beta} (s) =&~ h^3_{\si\rho\alpha\beta} 
\Bigl ( \pr_\mu^{\vphantom h} \pr_\nu^{\vphantom h} - {1\over d}\, \de_{\mu \nu}^{\vphantom h} \Bigl ){}
{1\over s^{d-2}} \\
H^4_{\mu\nu\si\rho\alpha\beta} (s) =&~\Bigl ( h^3_{\mu\nu\si\alpha}
\pr_\rho^{\vphantom h} \pr_\beta^{\vphantom h} + (\si\leftrightarrow \rho,
\alpha \leftrightarrow \beta)
- {2\over d}\, \de_{\si \rho}^{\vphantom h}
\big( h^3_{\mu\nu\lambda\alpha}\pr^\lambda
\pr_\beta^{\vphantom h}  + (\alpha\leftrightarrow \beta) \big) \nn\\
& -  {2\over d}\, \de_{\alpha \beta}^{\vphantom h}
\big( h^3_{\mu\nu\lambda\si}\pr^\lambda
\pr_\rho^{\vphantom h} + (\si\leftrightarrow \rho) \big) {}
+ {8\over d^2}\, \de_{\si\rho}^{\vphantom h} \de_{\alpha \beta}^{\vphantom h}
\Bigl ( \pr_\mu^{\vphantom h}
\pr_\nu^{\vphantom h} - {1\over d}\, \de_{\mu \nu}^{\vphantom h} \Bigl ){}
\Bigl ) {} {1\over s^{d-2}}
\end{align} 
and
\begin{align}
h^1_{\mu \nu}(\hs) =&~ \hs_\mu^{\vphantom h} \hs_\nu^{\vphantom h} - {1\over d} \, \de_{\mu \nu}^{\vphantom h} \, , \quad \hs_\mu^{\vphantom h} = {s_\mu^{\vphantom h} \over \sqrt {s^2}} \\
h^2_{\mu \nu \si \rho}(\hs) =&~ \hs_\mu^{\vphantom h} \hs_{\si}^{\vphantom h} \de_{\nu \rho}^{\vphantom h}
+(\mu \leftrightarrow \nu, \sigma \leftrightarrow \rho) -{4\over d} \hs_\mu^{\vphantom h} \hs_\nu^{\vphantom h}\de_{\si \rho}^{\vphantom h} -{4\over d} \hs_\si^{\vphantom h} \hs_\rho^{\vphantom h} \de_{\mu\nu}^{\vphantom h}
+ {4\over d^2} \de_{\mu \nu}^{\vphantom h} \de_{\si\rho}^{\vphantom h} \\
h^3_{\mu \nu \si \rho} =&~ \de_{\mu \si}^{\vphantom h} \de_{\nu\rho}^{\vphantom h}
+\de_{\mu \rho}^{\vphantom h} \de_{\nu\si}^{\vphantom h} -
{2\over d}\, \de_{\mu \nu}^{\vphantom h} \de_{\si\rho}^{\vphantom h} \\
h^4_{\mu\nu\si\rho\alpha\beta}(\hs) =&~ h^3_{\mu\nu\si\alpha}
\hs_\rho^{\vphantom h} \hs_\beta^{\vphantom h} + (\sigma \leftrightarrow \rho, \alpha   \leftrightarrow \beta) \nn\\
&  - {2\over d} \, \de_{\si \rho}^{\vphantom h}h^2_{\mu\nu\alpha\beta}(\hs)
- {2\over d} \, \de_{\alpha\beta}^{\vphantom h}h^2_{\mu\nu\si\rho}(\hs)
-{8\over d^2}\, \de_{\si\rho}^{\vphantom h}\de_{\alpha\beta}^{\vphantom h} h^1_{\mu\nu}(\hs) \\
h^5_{\mu\nu\si\rho\alpha\beta} =&~ \de_{\mu\si}^{\vphantom h}
\de_{\nu\alpha}^{\vphantom h}\de_{\rho\beta}^{\vphantom h}
 + (\mu \leftrightarrow \nu, \sigma \leftrightarrow \rho, \alpha   \leftrightarrow \beta)\nn\\
&-{4\over d}\,\de_{\mu\nu}^{\vphantom h} h^3_{\si\rho\alpha\beta}
-{4\over d}\,\de_{\si\rho}^{\vphantom h} h^3_{\mu\nu\alpha\beta}
-{4\over d}\,\de_{\alpha\beta}^{\vphantom h} h^3_{\mu\nu\si\rho}
-{8\over d^2} \, \de_{\mu\nu}^{\vphantom h}\de_{\si\rho}^{\vphantom h}
\de_{\alpha\beta}^{\vphantom h} \ .
\end{align} 
Define
\begin{align}
B_{\mu\nu\si\rho\alpha\beta\lambda}(s) =&~
A_{\mu\nu\si\rho\alpha\beta}(s) s_\lambda \nn\\
&+ {1\over (d+2)(d-1)} \Bigl (
(d+1)\B_{\mu\nu\si\rho\alpha\beta\lambda}(s) -
\B_{\mu\nu\si\rho\lambda\beta\alpha}(s) -
\B_{\mu\nu\si\rho\alpha\lambda\beta}(s) \nn\\
& -  d \, \de_{\alpha\lambda}^{\vphantom h}
A_{\mu\nu\si\rho\gamma\beta}(s) s^\gamma -  d \,\de_{\beta\lambda}^{\vphantom h}
A_{\mu\nu\si\rho\alpha\gamma}(s) s^\gamma + 2\de_{\alpha\beta}^{\vphantom h}
A_{\mu\nu\si\rho\lambda\gamma}(s) s^\gamma \Bigl ) 
\end{align} 
where 
\begin{align}
\B_{\mu\nu\si\rho\alpha\beta\lambda}(s) &= \half s^2 \pr_\lambda
A_{\mu\nu\si\rho\alpha\beta}(s) + s_\si^{\vphantom h} A_{\mu\nu\lambda\rho\alpha\beta}(s)
- \de_{\si\lambda}^{\vphantom h}\,  s^{\si'} A_{\mu\nu\si'\rho\alpha\beta}(s) \nn\\
&~~~ + s_\rho^{\vphantom h} A_{\mu\nu\si\lambda\alpha\beta}(s)- \de_{\rho\lambda}^{\vphantom h}\,  s^{\rho'} A_{\mu\nu\si\rho'\alpha\beta}(s) \\
A_{\mu\nu\si\rho\alpha\beta} (s)C_T &= {1\over s^d}   
\Big( a
h^5_{\mu\nu\si\rho\alpha\beta} + b
h^4_{\alpha\beta\mu\nu\si\rho}(\hs) + b'\big(
h^4_{\mu\nu\si\rho\alpha\beta}(\hs) +
h^4_{\si\rho\mu\nu\alpha\beta}(\hs)\big) \nn\\
&~~~ + c\, h^3_{\mu\nu\si\rho}h^1_{\alpha\beta}(\hs) + c'\big(
h^3_{\si\rho\alpha\beta}h^1_{\mu\nu}(\hs) +
h^3_{\mu\nu\alpha\beta}h^1_{\si\rho}(\hs)\big) \nn\\
&~~~ + e\, h^2_{\mu\nu\si\rho}(\hs)h^1_{\alpha\beta}(\hs) + e'\big(
h^2_{\si\rho\alpha\beta}(\hs)h^1_{\mu\nu}(\hs) +
h^2_{\mu\nu\alpha\beta}(\hs)h^1_{\si\rho}(\hs)\big) \nn\\
&~~~ +f\, h^1_{\mu\nu}(\hs)h^1_{\si\rho}(\hs)h^1_{\alpha\beta}(\hs) \Big) \ . 
\end{align}
Consistency conditions (i.e. the conservation and tracelessness) impose relations:\footnote{~We remark that \cite{Osborn:1993cr} misses a factor of $S_d$ in the equation relating $D$ to $C_T$.}
\begin{align}
 d\, D &={ C_T\over S_d}  \\
- {(d-2)(2a+b)-dc\over d(d+2)} + C &= 0 \\
 {2\over d^2}(d-1)\bigl ( (d-2)a-c\bigl ){} - {4\over d}\, C + D &=  {C_T \over 2 S_d} \\
{1\over d} \bigl ( 2(d-2)a-b-c\bigl ){}  + C &= {C_T \over 2 S_d}  \\
 {(d-2)(d+3)a -2b - (d+1)c \over d(d+2) } &= {C_T \over 4 S_d} \ . 
\end{align}
Also, $ b+b'=-2a, c'=c , e+e'=-4b'-2c$, $d^2 a + 2(b+b') -(d-2)b'-dc +e'= 0$, and $d(d+2)(2b'+c)+4(e+e')+f=0$.  One can write 
\begin{align}
f&=(d+4)(d-2)(4a+2b-c) \\
e'&=-(d+4)(d-2)a-(d-2)b +dc  \\
e &=(d+2)(da+b-c) \ .
\end{align}
There are three undetermined coefficients $a,b,c$.  A free scalar has 
\begin{align}
a=   {d^3\over 8(d - 1)^3 S_d^3} \ , ~~ b= -{ d^4\over 8 (d - 1)^3 S_d^3} \ , ~~ c=- { d^2 (d - 2)^2 \over 8 (d - 1)^3 S_d^3}  \ . 
\end{align}
For a free fermion,  $a=0,  b= -{ 2^{d\over 2} d^2 \over  16 S_d^3}, c=2b$. For a $d=4$ $U(1)$ field, $a=- {16\over  S^3_4}, b=0, c=4a$. 


\bibliographystyle{utphys}
\bibliography{4dOPEbib}

\end{document}